\documentclass[conference]{IEEEtran}
\usepackage[T1]{fontenc}
\usepackage{graphicx}
\usepackage{booktabs}
\usepackage{balance}
\usepackage{float}
\usepackage{xurl}
\usepackage{amsmath}
\usepackage[colorlinks=true,allcolors=black]{hyperref}
\usepackage{doi}
\usepackage{orcidlink}
\usepackage{cleveref}
\usepackage{cite}
\usepackage[font=footnotesize,labelfont=bf,labelsep=period]{caption}
\Crefname{figure}{Fig.}{Figs.}
\begin{document}

\title{Repository-Scale Performance Characterization of the IO500 Benchmark}
\author{
\IEEEauthorblockN{Aasish Kumar Sharma\,\orcidlink{0000-0002-7514-2340},
Anila Ghazanfar\,\orcidlink{0009-0003-2894-6893},
Sepehr Mahmoodianhamedani\,\orcidlink{0009-0000-0569-8047},
Sascha Safenreider\,\orcidlink{0009-0005-2287-0060},
Julian Kunkel\,\orcidlink{0000-0002-6915-1179}}
\IEEEauthorblockA{University of Göttingen, GWDG, Germany\\
\{aasish-kumar.sharma, anila.ghazanfar, s.mahmoodian, sascha.safenreider, julian.kunkel\}@gwdg.de}
}
\maketitle

\begin{abstract}
The IO500 benchmark provides a common basis for evaluating high-performance computing (HPC) storage systems, while its growing submission repository also offers an opportunity to study performance behavior across systems and time. This work presents a repository-scale characterization of 294 post-reset IO500 submissions from 116 sites spanning 2019--2025. We combine descriptive and temporal analysis with phase-level correlation, scale-sensitivity analysis, and benchmark-log examination to investigate how performance characteristics represented by individual benchmark phases relate to composite rankings. The results show strong relationships among phases measuring similar I/O behavior, but weaker and scale-sensitive relationships between bandwidth and metadata performance. Temporal and log-level analyses further reveal information that is not apparent from aggregate scores alone. These findings demonstrate the value of analyzing IO500 results beyond leaderboard rankings and identify important considerations concerning system scale, benchmark execution, and repository provenance when using community benchmark data for comparative and longitudinal HPC performance analysis.
\end{abstract}

\section{Introduction}
Storage performance is a critical component of HPC system performance, yet it is difficult to summarize because bandwidth, metadata throughput, access pattern, scale, and software configuration interact. IO500~\cite{kunkel2016establishing} addresses this problem by combining IOR~\cite{shan2007using}, MDTest~\cite{mdtest2007}, and parallel find into a standardized benchmark and public ranking. The resulting submission repository is more than a leaderboard: it is a multi-year corpus of storage measurements across architectures and sites. However, prior work has primarily evaluated individual systems or smaller subsets. A previous analysis of 61 IO500 submissions~\cite{kunkel2024ixpug} established useful phase relationships, but its two-year window could not characterize repository-wide temporal behavior.

We analyze the post-reset repository at substantially larger scale: 294 submissions from 116 sites over 2019--2025. We ask three questions: \emph{(RQ1)} how heterogeneous are composite and phase-level results, and how have scores changed over time; \emph{(RQ2)} how strongly are the 12 benchmark phases related, including after a scale-sensitivity check; and \emph{(RQ3)} what execution behavior is visible in runtime and stonewall records but hidden by composite scores? Our contribution is a repository-scale performance characterization that (i) defines an auditable post-reset IO500 population, (ii) quantifies temporal and phase-level behavior with effect sizes and FDR-controlled correlations, (iii) tests the scale sensitivity of cross-domain relationships, and (iv) links ranking-level results to runtime and stonewall effects. The study therefore evaluates both what IO500 captures and what its composite ranking obscures.

\section{Related Work}
IOR~\cite{shan2007using} and MDTest~\cite{mdtest2007} are established HPC storage benchmarks, while IO500~\cite{kunkel2016establishing} combines multiple access patterns into a common ranking methodology. Prior studies have examined I/O variability~\cite{lofstead2010managing}, continuous workload characterization~\cite{carns2011understanding,snyder2016modular}, and individual parallel storage technologies such as Lustre~\cite{schwan2003lustre}, GPFS~\cite{schmuck2002gpfs}, and DAOS~\cite{daos2024,hennecke2023understanding}. Monnier et al.~\cite{monnier2019profiling} combined IO500 with profiling for platform-level evaluation. The closest repository-oriented work~\cite{kunkel2024ixpug} analyzed 61 submissions from 2021--2022. Our study differs by defining a post-reset population from the official SC19 boundary, analyzing nearly five times as many submissions over a longer period, testing scale sensitivity, and incorporating runtime/stonewall records. The objective is not to predict acquisition outcomes, but to establish what can and cannot be inferred statistically from the public benchmark corpus.

\section{Benchmark and Methodology}
\label{sec:method}
IO500 reports four IOR bandwidth measurements (easy/hard write/read), seven MDTest metadata measurements, and parallel find, yielding 12 phase-level metrics. Nine contribute to the composite scores. The bandwidth score is the geometric mean of four IOR results; the metadata score combines MDTest easy-create/stat, hard-create/stat, and find; the overall score is the geometric mean of bandwidth and metadata scores. Write/create phases use a nominal 300-s stonewall followed by wear-down, while subsequent phases operate on the generated data.

We parsed 365 packages from the public submission repository~\cite{io500submission2025}. Directory year alone cannot identify the post-reset population because the 2019 directory contains historical results from earlier lists. The SC19 benchmark revision explicitly made some former results non-comparable; therefore, we use the official SC19 Full List~\cite{io500sc19} as the reset boundary. A 2019 package is retained only when its published overall, bandwidth, and metadata score triple matches an SC19 Full-List entry to two-decimal precision. This retains 38 of 105 year-tagged 2019 packages; 67 legacy 2019 and four pre-2019 packages are excluded. Records from 2020--2025 are retained, producing 294 submissions from 116 sites. Complete phase data are available for 222 submissions.

Because score distributions are strongly skewed, we use nonparametric statistics. Pairwise phase relationships use Spearman's $r_s$ over 222 complete records; Benjamini--Hochberg (BH) correction~\cite{benjamini1995controlling} controls the false discovery rate over 66 comparisons at $\alpha=0.05$. To assess scale sensitivity without conflating processes with nodes, we repeat the cross-domain analysis after dividing phase scores by recorded total process count for the 116 complete records containing that field. Temporal differences compare 2019--2022 ($n=148$) with 2023--2025 ($n=146$) using the Mann--Whitney $U$ test~\cite{mannwhitney1947} and rank-biserial effect size $r_{rb}$. Finally, runtime and stonewall fields are analyzed directly from result logs.

Four limitations constrain interpretation. First, client-node counts are unavailable in the parsed packages, so per-process normalization is only a sensitivity test. Second, 59 of 116 sites submit more than once, violating strict independence in the temporal test. Third, later IO500 version differences remain after the explicit SC19 reset. Fourth, participation is voluntary; trends describe the repository population, not all HPC storage systems.

\textit{Data validation:} We retain the complete parsed snapshot separately from the analysis population so that reset filtering is auditable. For every retained record, we verify the published composite identity $S_{overall}\approx\sqrt{S_{BW}S_{MD}}$; no analyzed record differs by more than 0.1\%. Filesystem labels are standardized into nine named families plus ``Other'' (\Cref{tab:fs}). The latter contains 102 of 294 records and is intentionally retained rather than discarded; the correlation robustness test below evaluates whether this heterogeneous category changes the principal result. These checks address data provenance without silently repairing benchmark measurements.

\begin{table}[t]
\caption{Largest filesystem groups in the post-reset population.}
\label{tab:fs}
\centering
\footnotesize
\begin{tabular}{lrr@{\hspace{5mm}}lrr}
\toprule
\textbf{FS} & \textbf{$n$} & \textbf{\%} & \textbf{FS} & \textbf{$n$} & \textbf{\%}\\
\midrule
Lustre & 66 & 22.4 & DAOS & 40 & 13.6\\
GPFS/SS & 22 & 7.5 & WekaFS & 18 & 6.1\\
VAST & 11 & 3.7 & BeeGFS & 11 & 3.7\\
GekkoFS & 10 & 3.4 & CephFS & 9 & 3.1\\
IME & 5 & 1.7 & Other & 102 & 34.7\\
\bottomrule
\end{tabular}
\end{table}

\section{Results}
\label{sec:results}
\subsection{Distributions and Temporal Change}
The 294 overall scores range from 0.58 to 210,255, with median 119.4 and coefficient of variation (CV) 7.40. Bandwidth has median 38.9~GiB/s (CV 4.26), whereas metadata has median 318.7~kIOPS (CV 9.27). Across the 222 complete records, metadata-phase CVs (7.02--11.34) exceed IOR-phase CVs (2.52--4.64), with find the most variable phase (CV 11.34; \Cref{tab:phase}). Thus, a single composite rank summarizes a population whose metadata behavior is markedly more dispersed than its bandwidth behavior.

\begin{table}[t]
\caption{Phase-level medians and coefficients of variation ($n=222$).}
\label{tab:phase}
\centering
\scriptsize
\begin{tabular}{lrr@{\hspace{4mm}}lrr}
\toprule
\textbf{Phase} & \textbf{Median} & \textbf{CV} & \textbf{Phase} & \textbf{Median} & \textbf{CV}\\
\midrule
IOR easy-w & 106 & 3.97 & MD easy-w & 387 & 8.24\\
IOR easy-r & 126 & 2.52 & MD easy-s & 1,213 & 7.86\\
IOR hard-w & 12 & 4.64 & MD easy-d & 323 & 8.24\\
IOR hard-r & 41 & 3.89 & MD hard-w & 74 & 7.71\\
Find & 2,642 & 11.34 & MD hard-r & 379 & 7.02\\
 & & & MD hard-s & 832 & 7.87\\
 & & & MD hard-d & 80 & 8.56\\
\bottomrule
\end{tabular}
\begin{flushleft}\scriptsize IOR medians are GiB/s; MDTest and Find medians are kIOPS.\end{flushleft}
\end{table}

\begin{table}[t]
\caption{Temporal comparison of post-reset composite scores.}
\label{tab:temporal}
\centering
\footnotesize
\begin{tabular}{lrrrr}
\toprule
\textbf{Metric} & \textbf{2019--22} & \textbf{2023--25} & \textbf{Ratio} & \textbf{$r_{rb}$}\\
 & \textbf{median} & \textbf{median} & & \\
\midrule
Overall & 77.1 & 179.2 & 2.32$\times$ & 0.25\\
Bandwidth & 28.3 & 73.7 & 2.60$\times$ & 0.26\\
Metadata & 166.4 & 478.0 & 2.87$\times$ & 0.24\\
\bottomrule
\end{tabular}
\vspace{-1mm}
\begin{flushleft}\scriptsize All Mann--Whitney comparisons have $p<0.001$; repeated sites mean these $p$-values should be interpreted cautiously.\end{flushleft}
\end{table}

\Cref{tab:temporal} shows higher recent-period medians for all composite metrics. The overall median is 2.32$\times$ higher, while bandwidth and metadata are 2.60$\times$ and 2.87$\times$ higher. These are population-level differences rather than a monotonic growth law: yearly medians are not monotonic, repeated sites weaken independence, and the filesystem mix changes between periods. Phase-level change is also uneven: IOR-easy read increases 2.3$\times$ ($p=0.004$), whereas IOR-hard write shows no significant difference (1.4$\times$, $p=0.94$); MDTest-hard stat increases 2.7$\times$ ($p=0.033$), whereas MDTest-easy stat does not (0.8$\times$, $p=0.78$). Aggregate score change therefore masks access-pattern-specific behavior.

\subsection{Phase Correlations and Scale Sensitivity}
\begin{figure}[t]
\centering
\includegraphics[width=0.94\columnwidth]{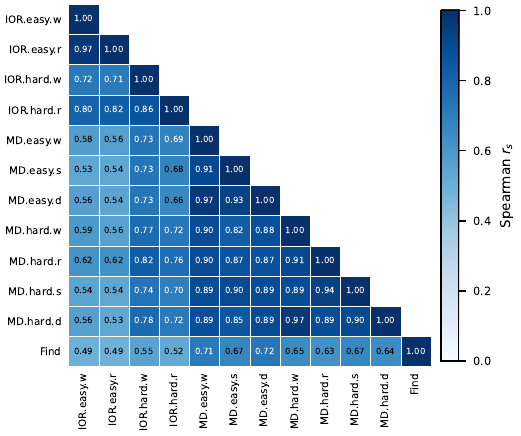}
\caption{Spearman correlations among 12 phases ($n=222$). The block structure separates stronger within-domain from generally weaker cross-domain associations.}
\label{fig:corrplot}
\end{figure}

\Cref{fig:corrplot} shows a clear block structure. IOR-IOR correlations range from $r_s=0.71$ to 0.97 and MDTest-MDTest from 0.82 to 0.97, whereas IOR-to-metadata-domain correlations are lower at 0.49--0.82. All 66 pairwise correlations remain significant after BH correction. Excluding the 71 ``Other'' filesystem records from the phase subset leaves the cross-domain range similar (0.43--0.76), so heterogeneous naming does not drive the main pattern. Relative to the earlier 61-submission study~\cite{kunkel2024ixpug}, which reported cross-domain values of 0.70--0.95, the larger corpus supports a more differentiated view: strong association within domains does not imply that bandwidth and metadata phases are redundant.

Raw scores are scale-sensitive. For the 116 complete submissions with recorded process count, raw cross-domain correlations are $r_s=0.52$--$0.87$; after dividing each phase score by process count, the range falls to $0.28$--$0.81$. This does not constitute node normalization, but it demonstrates that common scale contributes to raw phase association. It also strengthens the case for richer machine-readable system-scale metadata in future IO500 submissions.

\subsection{Log-Derived Execution Effects}
All 222 complete write/create records exceed the nominal 300-s stonewall, with median elapsed times of approximately 331--364~s after wear-down. In contrast, 49 read/stat/find observations finish in under 10~s. Such short executions are consistent with cache or client-side data-retention effects, although the logs cannot identify the mechanism.

\begin{figure}[t]
\centering
\includegraphics[width=0.94\columnwidth]{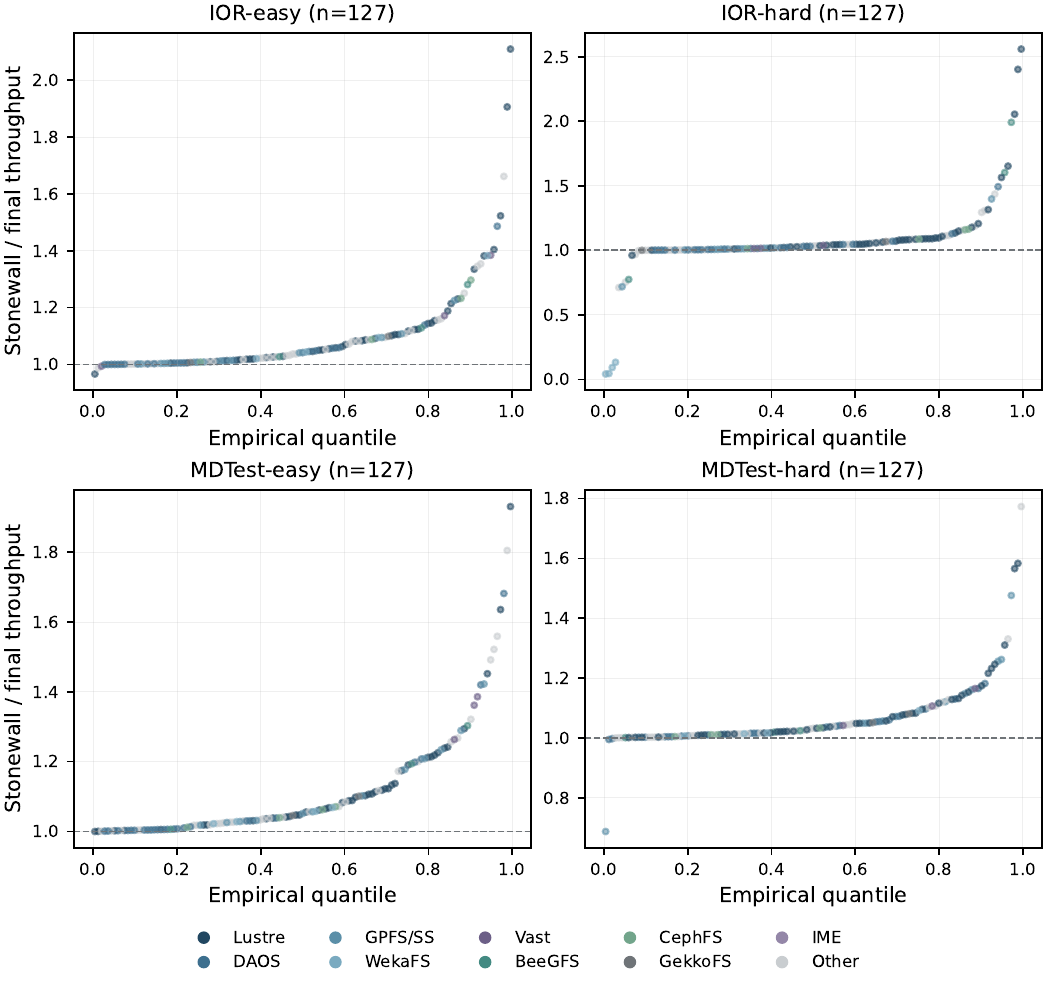}
\caption{Empirical quantiles of stonewall-to-final throughput ratios ($n=127$ submissions with stonewall cutoff records). Upper tails reveal wear-down losses that are hidden by the composite score.}
\label{fig:stonewall}
\end{figure}

For stonewalled phases, we compare throughput at the cutoff with final scored throughput after wear-down for the 127 submissions whose logs contain a stonewall cutoff record (\Cref{fig:stonewall}); the remaining 95 complete records do not include this field. Median ratios are only 1.03--1.05, indicating limited typical loss, but the upper tail is material: IOR-hard reaches a 95th percentile of 1.54 and maximum of 2.56; MDTest-hard reaches 1.77. Hence stragglers can substantially reduce the final score for a minority of submissions even when the median effect is small. These execution-level effects are invisible in the overall IO500 rank.

\section{Discussion and Conclusion}
\label{sec:discussion}
The results support four conclusions relevant to benchmark interpretation and performance analysis (\Cref{tab:implications}). First, IO500's overall score is effective for ranking but insufficient for workload-specific interpretation. Strong within-domain and weaker cross-domain correlations show that bandwidth and metadata capture complementary dimensions. Two post-reset submissions illustrate this directly: an EPCC GekkoFS result (152.3 overall) reports 24.8~GiB/s bandwidth and 934.1~kIOPS metadata, whereas a JCAHPC IME result (146.3 overall) reports 341.3~GiB/s and 62.7~kIOPS. Similar overall scores can therefore represent markedly different I/O profiles.

\begin{figure*}[t]
\centering
\begin{minipage}[t]{0.48\textwidth}
\centering
\includegraphics[width=\linewidth]{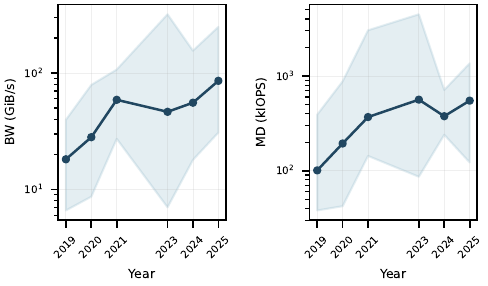}
\textbf{(a)} Yearly bandwidth and metadata medians with interquartile ranges.
\end{minipage}\hfill
\begin{minipage}[t]{0.48\textwidth}
\centering
\includegraphics[width=\linewidth]{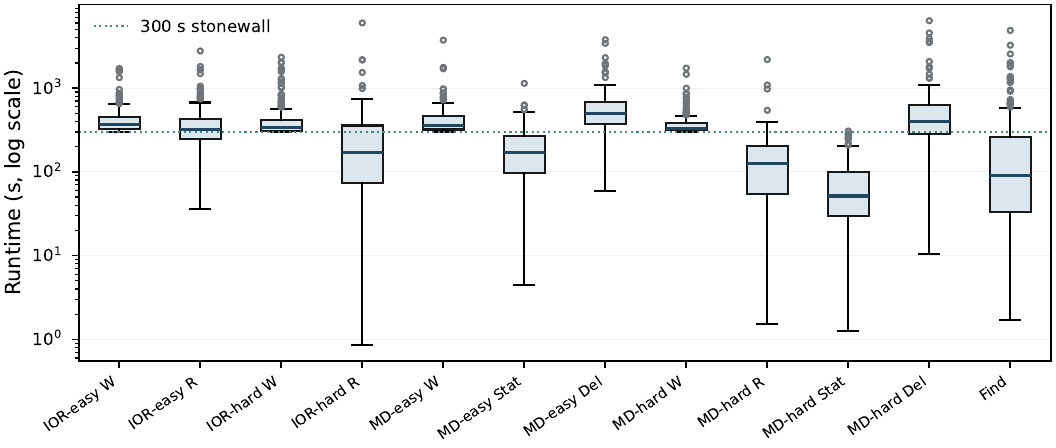}
\textbf{(b)} Phase runtime distributions; the dashed line marks the 300-s stonewall.
\end{minipage}
\caption{Complementary longitudinal and execution-level views of IO500 performance. Both yearly score series are non-monotonic; 2022 is absent from panel~(a) because no 2022 submission appears in the complete-phase population ($n=222$). Runtime distributions in panel~(b) expose behavior not represented in the composite ranking.}
\label{fig:temporal_runtime}
\end{figure*}

Second, the 2.3$\times$ difference in overall median between the two periods is a repository-level distributional shift, not a causal estimate of hardware progress (\Cref{fig:temporal_runtime}a). Yearly medians are non-monotonic, 59 of 116 sites submit repeatedly, the participant and filesystem mix changes, and benchmark versions evolve. The phase-level results reinforce this caution: some phases change substantially while others do not.

Third, raw phase correlations are partly scale-driven. On the 116 records with process counts, per-process normalization reduces cross-domain $r_s$ from 0.52--0.87 to 0.28--0.81. Process count is not client-node count, so this is deliberately a sensitivity test rather than a normalized benchmark. Nevertheless, it demonstrates why future IO500 submissions would benefit from stable machine-readable client-node, process-placement, benchmark-version, and storage/interconnect metadata.

Fourth, log-level execution behavior adds information that rankings discard. Median stonewall-to-final ratios are close to one, yet the upper tail reaches 2.56 for IOR-hard; 49 read/stat/find observations complete in under 10~s. The latter are consistent with cache or client-side retention effects but do not identify a mechanism. Thus, runtime and wear-down records are useful diagnostics when systems achieve similar composite scores through different execution behavior.

\begin{table*}[t]
\caption{Performance-analysis implications of the repository-scale findings.}
\label{tab:implications}
\centering
\footnotesize
\begin{tabular}{p{0.22\textwidth}p{0.32\textwidth}p{0.38\textwidth}}
\toprule
\textbf{Observation} & \textbf{Evidence} & \textbf{Implication for benchmark use/design}\\
\midrule
Composite-score ambiguity & Similar overall scores can pair high bandwidth with low metadata performance, or vice versa. & Use phase/domain scores for workload-oriented interpretation; retain the composite score for ranking.\\
Scale-sensitive association & Cross-domain $r_s$ falls from 0.52--0.87 to 0.28--0.81 after per-process scaling on the same 116 records. & Record machine-readable node/process placement to enable defensible normalization.\\
Nonuniform longitudinal change & Period medians rise, but yearly and phase-level changes are non-monotonic. & Preserve benchmark-version and stable system/site identifiers for longitudinal analysis.\\
Execution effects & Stonewall wear-down and short read/stat runs are not represented by final rankings. & Preserve runtime/stonewall metadata as first-class benchmark evidence.\\
\bottomrule
\end{tabular}
\end{table*}

\textit{Benchmark-design implications:} The results suggest three concrete directions for performance-benchmark repositories. First, composite scores should remain accompanied by phase-level results because the correlation structure shows strong within-domain redundancy but materially weaker bandwidth--metadata association; this enables workload-oriented interpretation without changing the ranking definition. Second, longitudinal analysis requires explicit machine-readable provenance. Stable identifiers for benchmark version, site/system, client-node count, process placement, and storage/interconnect configuration would permit per-node normalization and repeated-measures analyses that are not defensible from the present repository alone. Third, execution metadata should be treated as part of the benchmark evidence rather than ancillary logs: stonewall and runtime fields expose wear-down and short-execution behavior that cannot be reconstructed from the final composite score. These additions would improve the repository as a performance-analysis resource while preserving IO500's existing benchmark phases and public ranking.

\textit{Validity and reproducibility:} The principal threats are repository self-selection, repeated sites, benchmark-version evolution, and incomplete system-scale metadata. We therefore report effect sizes with $p$-values, use the SC19 list as an explicit reset boundary, and avoid causal or node-normalized claims unsupported by the repository. The analysis pipeline preserves the parsed snapshot, applies population filtering deterministically, verifies $S_{overall}\approx\sqrt{S_{BW}S_{MD}}$, and regenerates statistics and figures from the filtered dataset.

In summary, 294 post-reset IO500 submissions from 116 sites establish a reproducible picture of extreme score heterogeneity, higher recent-period medians with nonuniform phase change, distinct but scale-sensitive bandwidth/metadata relationships, and execution effects hidden by rankings. The IO500 repository is therefore valuable not only as a leaderboard but as an HPC performance-analysis dataset, provided its statistical and metadata limitations are explicit.

\section*{Acknowledgment}
The authors thank the IO500 community and the IO500 foundation (io500.org), and the VI4IO initiative for making the submission data publicly available. The authors also thank Dr. Jay Lofstead for his constructive feedback.

\bibliographystyle{IEEEtran}
\bibliography{references}
\end{document}